\documentstyle[psfig,referee]{l-aa-mpa}

\newif\ifAMStwofonts
\AMStwofontstrue

\newcommand{\dm}{\mbox{$m-M$}}
\newcommand{\sigmadm}{\mbox{$\sigma_{m\!-\!M}$}}
\newcommand{\deltadm}{\mbox{$\delta_{m\!-\!M}$}}
\newcommand{\mv}{\mbox{$M_{V}$}}
\newcommand{\mi}{\mbox{$M_{I}$}}
\newcommand{\mbol}{\mbox{$M_{\rm bol}$}}
\newcommand{\imax}{\mbox{$I^{\rm max}$}}
\newcommand{\iomax}{\mbox{$I_0^{\rm max}$}}
\newcommand{\mimax}{\mbox{$M_{I}^{\rm max}$}}

\newcommand{\bv}{\mbox{$B-V$}}

\newcommand{\vi}{\mbox{$V-I$}}

\newcommand{\dmo}{\mbox{$(m-M)_{0}$}}

\newcommand{\evi}{\mbox{$E_{V-I}$}}
\newcommand{\feh}{\mbox{[Fe/H]}}
\newcommand{\logt}{\mbox{$\log(t/{\rm yr})$}}
\newcommand{\Msun}{\mbox{$M_{\odot}$}}

\newcommand{\sub}[1]{\mbox{$_{\rm #1}$}}

\newcommand{\Mto}{\mbox{$M\sub{TO}$}}
\newcommand{\Mhef}{\mbox{$M\sub{Hef}$}}
\newcommand{\Mhe}{\mbox{$M\sub{cl}$}}

\newcommand{\logTe}{\mbox{$\log T\sub{eff}$}}
\newcommand{\logL}{\mbox{$\log(L/L_{\odot})$}}
\newcommand{\diff}{\mbox{d}}
\newcommand{\beq}{\begin{equation}}
\newcommand{\eeq}{\end{equation}}
\newcommand{\beqa}{\begin{eqnarray}}
\newcommand{\eeqa}{\end{eqnarray}}
\newcommand{\benu}{\begin{enumerate}}
\newcommand{\eenu}{\end{enumerate}}
\newcommand{\bite}{\begin{itemize}}
\newcommand{\eite}{\end{itemize}}
\newcommand{\bdes}{\begin{description}}
\newcommand{\edes}{\end{description}}
\newcommand{\refeq}[1]{equation (\protect\ref{#1})}
\newcommand{\reffig}[1]{Fig.\ \protect\ref{#1}}

\newcommand{\refsec}[1]{Section \protect\ref{#1}}
\newcommand{\comment}[1]{}

\begin{document}

\title{Fine structure of the red giant clump from {\em Hipparcos}
	data, and distance determinations based on its mean
	magnitude\thanks{Based on data from the ESA {\em Hipparcos} 
	astrometry satellite.}}
\author{L\'eo Girardi\inst{1} \and Martin A.\ T.\ Groenewegen\inst{1}
	\and Achim Weiss\inst{1} \and Maurizio Salaris\inst{2,1}}
\institute{Max-Planck-Institut f\"ur Astrophysik, 
	Karl-Schwarzschild-Stra{\ss}e 1, D-87540 Garching bei M\"unchen,
	Germany \and 
	Astrophysics Research Institute, Liverpool John Moores
	University,
	Byrom Street, Liverpool L3 3AF, UK
	} 
\date{Accepted 19?? ???.
      Received 1998 ???;
      in original form 1998 ???}

\maketitle
\markboth{L.~Girardi et al.: Fine structure of the red giant clump}
{L.~Girardi et al.: Fine structure of the red giant clump}

\label{firstpage}
\thispagestyle{empty}
\mbox{\ }\clearpage\setcounter{page}{1}

\thispagestyle{empty}
\parbox[b]{15cm}{\mbox{~~~~} \\[4.0cm] 
\begin{abstract}
The $I$-band brightness $M_I$ of clump stars is
a possible distance indicator for stellar populations. Investigations
have shown that $M_I$ is almost insensitive to the $(V-I)$ colour
within the clump. Based on this, it was assumed that $M_I$ was
insensitive to age and composition of the stellar population and
therefore an ideal standard candle, which could be calibrated with
local clump stars, whose absolute brightness is known from
{\em Hipparcos} parallaxes. This resulted in a distance to the LMC
about 15\% shorter than usually determined.

In the present paper we show that with a population synthesis approach we can
reproduce the constancy of $M_I$ with colour for the local {\sl
Hipparcos} clump sample. Nevertheless, $M_I$ is not a
constant among different populations, but depends on metallicity. As a
result, the determined distance modulus to the LMC of $18.28\pm0.18$
mag is in
better agreement with standard values. This resolves, at
least partially, the controversial result obtained by the assumption
of a universal value for $M_I$. 

Particularly
remarkable is our prediction that stars slightly heavier than the
maximum mass for developing degenerate He cores, \Mhef, should define
a secondary, clumpy structure, about 0.3~mag below the bluest
extremity of the red clump. Both features are well separated in the \mi\
vs.\ \vi\ diagram of metal-rich stellar populations. Indeed, this
secondary clump can be clearly identified in the {\em Hipparcos}
database of stars with reliable $I$ photometry and parallax errors
smaller than 10\%.  Since the stars in this feature should represent a
narrow range of masses, their mass determination, e.g.\ by the use of
binary systems, can provide information about the efficiency of
convective overshooting from stellar cores.  

Our investigation demonstrates that the RGB clump cannot be used as a 
distance indicator without proper knowledge and modelling of the
population under investigation. In addition, there remain unsolved
problems in the models, such as correct bolometric corrections and
colour transformations. \vspace{2.0cm}

\comment{stars:horizontal branch --} 
\comment{stars: luminosity function, mass function --}
solar neighbourhood -- 
Galaxy: stellar content -- 
galaxies: distances and redshifts -- 
Magellanic Clouds 
\end{abstract} 
}

\thispagestyle{empty}
\mbox{\ }\clearpage\setcounter{page}{1}

\section{Introduction}
\label{sec_intro}

The red giant clump is an easily recognizable feature in many 
colour-magnitude diagrams (CMD). It consists of stars of rather low mass,
which are currently undergoing their central helium
burning. Physically, it is identical to the horizontal branch in
globular clusters made up
by less massive and more metal-poor stars. 

Paczy\'nski \& Stanek (1998) recently determined the
mean absolute brightness ($M_I$) of local clump stars making use of
{\em Hipparcos} parallaxes. They selected the stars from the {\em Hipparcos}
catalogue (ESA 1997) with parallax measured to better than 10\%. Then they
determined the mean \mi\ magnitude of the clump stars,
$\mimax=-0.185\pm0.016$, by fitting a gaussian-like curve to the
magnitude distribution of 657 stars inside the box defined by
$0.8<(\vi)<1.25$, $1.1>\mi>-1.4$. Selecting stars in two colour
subintervals, $0.8<(\vi)<1.0$ and $1.0<(\vi)<1.25$, they found values
of \mimax\ formally indistinguishable from the mean. The same
independence of the apparent $I$-band brightness \imax\ in the \vi\
colour was also found for the clump 
stars in Baade's Window, over the entire $0.8<(\vi)<1.4$ interval.
Based on these results, Paczy\'nski \& Stanek (1998) assumed \mimax\
to be independent of the properties of the observed stellar
populations, at least for stars with $0.8<(\vi)<1.25$. Under this
assumption, the galactocentric distance was 
determined by comparing the apparent $I$ magnitude of Baade's Window
clump stars, \imax, with the reference value obtained from the {\em
Hipparcos} sample.  The same process was repeated for stars in M31
(Stanek \& Garnavich 1998), the Magellanic Clouds (Udalski et al.\
1998), and the LMC alone (Stanek, Zaritski \& Harris 1998). In all
these cases, the mean \imax\ value was found to be nearly independent
of the \vi\ colour sampled.

For both the bulge and M31, the distances obtained were essentially in
agreement with those 
obtained from other methods. In the case of the Magellanic Clouds,
however, distances turned out to be significantly shorter than those
derived by Cepheid stars and commonly accepted: 
Udalski et al.\ (1998) and Stanek, Zaritsky \&
Harris (1998) find that the relatively well-settled distance modulus
of the LMC of about $\dmo=18.5$~mag could be overestimated by 0.45~mag, or a
factor of about 15\% in distance, with respect to the real one. Since
clump stars provide a one-step distance, this
lead them to claim that {\em other} distance indicators (such as
Cepheids) were erroneous and should be re-investigated.

Since Magellanic Cloud stars have mean metallicities well below
those of local stars and of stars that define the clump in the bulge
and in M31 over the $0.8<(\vi)<1.25$ interval, the suspicion that
metallicity effects may be causing the discrepant results, is
obvious. 
According to the above authors, theoretical models show weak
dependence of \mbol\ and \mi\ (for clump stars) on either age or
chemical composition. That is a surprising statement, since models in
the literature (e.g.\ Sweigart \& Gross 1976; Seidel, Demarque \&
Weinberg 1987; Bertelli et al.\ 1994, and references therein) show
that clump stars of different masses and metallicities may have
luminosities differing of up to 0.5~mag (see 
\refsec{sec_stars}).  The main support for using \mimax\ as a
standard candle comes, instead, from the observed independence of
\mimax\ on \vi\ (and hence supposedly metallicity).

Cole (1998) accordingly proposed a revision of the clump distance to
the LMC, based on the mean age and
metallicity differences between the LMC and local stars. Considering
these differences, and making use of the (theoretical) dependence of
clump magnitude on both parameters, he shows that the LMC red clump
should be about 0.32~mag brighter than the local disk one, and obtains
a distance modulus of $18.36\pm0.17$~mag to the LMC. Beaulieu \&
Sackett (1998) obtained a good model for the LMC clump by adopting a
distance modulus of $18.3$ from isochrone fitting.

However, this kind of first-order explanation seems to be in contrast
with the 
observation that \mimax\ is almost constant at different \vi\
colours in different stellar systems. Since the clump colour is
usually considered as indicative of metallicity, the \mimax\
constancy with colour strongly suggests that it is, in reality,
inpedendent of metallicity. The question arises if theoretical models
can explain this constancy in a composite stellar population, but
simultaneously predict brighter clumps at lower metallicities, as
required to obtain the usual LMC distance. That is one of the
questions we are going to address in the following.

In this work, we intend to examine the fundamentals of the red clump
method, with the aid of evolutionary models and isochrone calculations
which should represent, as far as possible, the standard theoretical
predictions for the behaviour of clump stars in stellar populations of
different ages and metallicities. In \refsec{sec_stars} we briefly
describe the theoretical stellar models and isochrones we used, and
the general predictions for the \mi\ magnitude and colours of the
clump stars; in \refsec{sec_clump} we show that the models predict a
fine structure of the red clump, which is indeed observed in the \mi\
vs.\ \vi\ diagram from {\em Hipparcos} data. In the light of these
results, in \refsec{sec_lf} we show how the mean \mi\ for a composite
stellar population can be, at the same time, almost independent of the
\vi\ colour sampled, and dependent on the mean metallicity of the
population observed (being in general brighter for lower
metallicities); this behaviour, together with our present
knowledge on the star formation history of the LMC, helps
to put the distance modulus of this galaxy closer to the more traditional
value of about 18.5~mag (\refsec{sec_lmc}).  Finally, in
\refsec{sec_comments} we comment on the accuracy in distance
determinations that can be obtained from the red clump method.

\section{Theoretical models for clump stars}
\label{sec_stars}

Stellar evolution theory predicts a fundamental dichotomy between the
evolutionary behaviour of stars which do, and those which do not,
develop electron degenerate cores after central hydrogen
exhaustion. Stars of masses lower than about 2~\Msun\ develop
degenerate cores, and climb the red giant branch (RGB) until the core
mass grows to about 0.45~\Msun. Then a relatively strong He-flash
event lifts degeneracy, and the star settles on the core He-burning
phase. The maximum mass of stars that follow this evolutionary scheme
is usually denoted as \Mhef\ (e.g.\ Renzini \& Buzzoni
1986; Maeder \& Meynet 1989; Chiosi, Bertelli \& Bressan 1992).  Stars
of masses slightly above \Mhef\ have a weakly-degenerate core, and are
able to ignite helium with a lower core mass, of about 0.33 \Msun;
therefore their RGB phase is significantly abbreviated.  For stars of
higher masses, the core mass at He ignition becomes an increasing
function of stellar mass, and the evolutionary phase equivalent to the
RGB is practically missing.  This difference in the core mass
reflects directly on the luminosity of the core He-burning stars: for
$M<\Mhef$ their luminosity is almost constant at $\logL\sim1.5$, which
gives origin to the clump of red giant stars (or to the horizontal
branch in the case of Pop.~II stars); for
masses slightly above \Mhef\ this luminosity is predicted to have a
minimum value, about 0.4~mag below the clump of lower mass stars; and
for still higher total mass the He burning stars shift progressively to higher
luminosities, so that they do not correspond to the `red clump'
anymore. The exact values of the stellar masses and luminosities
depend slightly on the assumed values of $Z$ and $Y$. The 
limiting mass \Mhef\ also depends on the extent of
convective cores during the main sequence (MS) phase: for solar
composition models and adopting the classical Schwarzschild criterion
for convective instability, one finds $\Mhef\simeq2.4$~\Msun\ 
(e.g.\ Sweigart et al.\
1990; Girardi et al.\ 1998), while values as low as 1.7~\Msun\ are
found when overshooting schemes are adopted (see e.g.\ Maeder \&
Meynet 1989; Bertelli, Bressan \& Chiosi 1985; Chiosi et al.\ 1992).

\subsection{The present stellar tracks}

We illustrate here the expected location of clump stars in both the
theoretical, \logL\ vs.\ \logTe, and observational, \mi\ vs.\ \vi,
planes, for a particular set of stellar models. 

We use a large set of evolutionary tracks
(Girardi \& Bertelli 1998; Girardi et al.\ 1998), with metallicities
ranging from $Z=0.001$ to $Z=0.03$ and with the helium content given
by the relation $Y=0.23+2.25\,Z$. This relation reproduces the initial solar
helium content of 0.273 (Girardi et al.\ 1996)
for the primordial value of $Y=0.230$
(Torres-Peimbert, Peimbert \& Fierro 1989; Olive \& Steigman 1995).
The typical
mass resolution for low-mass stars (i.e.\ those with $M<\Mhef$) is 
0.1~\Msun, increasing to 0.2 to 0.5~\Msun\ for intermediate-mass stars
(i.e.\ $M>\Mhef$). The $Z=0.019$ models, for instance, were computed
for masses 0.5, 0.55, 0.6, 0.7, 0.8, 0.9, 1, 1.1, 1.2, 1.3, 1.4, 1.5,
1.6, 1.7, 1.8, 1.9, 1.95, 2, 2.2, 2.5, 3, 3.5~\Msun. We used an
updated version of the Padova code (see Bertelli et al.\ 1994, and
references therein), with the input physics as described in Girardi et
al.\ (1996) and Girardi \& Bertelli (1998). We assume
moderate convective overshooting (see e.g.\ Chiosi et al.\ 
1992 for a review), so that we find $\Mhef=2.0$~\Msun\ for our
$Z=0.019$ tracks.

For $M<\Mhef$, the RGB evolutionary sequences are stopped at the
He-flash, and continued from a quiescent He-burning model with the
same core mass and chemical structure as the last RGB
model. Therefore, all the relevant evolutionary phases previous to the
onset of the thermally-pulsing regime of the asymptotic giant branch
(AGB) are included in the tracks. Models are computed at
constant mass, since the effect of mass loss can be realistically
considered during the preparation of isochrones (see
e.g.\ Renzini 1977; Bertelli et al.\ 1994).

\begin{figure}
\begin{center}
\psfig{file=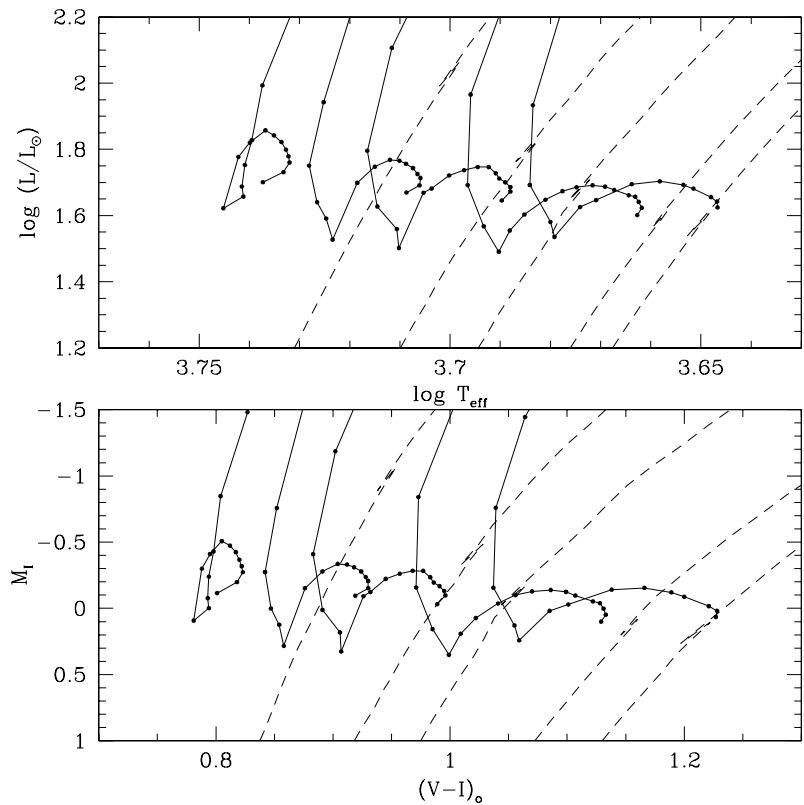,width=11cm}
        \caption[]{Position of the ZAHB (onset of quiescent
He-burning) for stellar evolutionary tracks of different masses and
metallicities, in the HR (top panel) and \mi\ vs.\ \vi\ (bottom panel)
diagrams. Dots represent the computed models, at typical mass
intervals of 0.1~\Msun\ for $M\la2$~\Msun, and 0.2 to 0.5~\Msun\ for
$M\ga2$~\Msun. For the sake of clarity, only models with $M>0.7$~\Msun\
are plotted. Metallicities are, from left to right, $Z=0.001$, 0.004,
0.008, 0.019 and 0.03. The dashed lines correspond to the RGBs of
4~Gyr isochrones with the same values of metallicity.}
\label{fig_zahb}
\end{center}
\end{figure}

The upper panel of \reffig{fig_zahb} shows the location of the
zero-age horizontal branch (ZAHB) for $M\ge0.7$~\Msun\ in the HR
diagram. We understand the ZAHB in this context as the location of
stars beginning their central helium burning phase, independent of
initial stellar mass.
It is essentially defined by the lowest luminosity
reached by our He-burning tracks, so that the depicted lines can be
thought as the lower envelope of clump stars in the HR diagram, as a
function of mass and metallicity.  ZAHB models with masses between 0.7
and $\Mhef$ describe a kind of hook or semi-circle in the
diagram, which is limited to a small range of luminosities [e.g.\
$1.5\la\logL\la1.7$ for $Z=0.019$]. It reflects the small range
of core masses for these stars at the moment of He-ignition. At the
hottest extremity of 
this hook, the models with $M>\Mhef$ depart to high luminosities:
these are the stellar models in which He burning started under
non-degenerate conditions.

The precise shape of the ZAHB in the HR diagram is determined by the
interplay of several factors, mainly related to the properties of the
stellar envelopes (e.g.\ its total mass, $Y$ and $Z$), and to the core
mass for those stars close and above \Mhef. The general behaviour is that
stars with lower masses ($M\la1$~\Msun) or with $M\simeq\Mhef$
have lower luminosities than stars with $M\simeq1.4$~\Msun, and  that
in this mass range higher masses correspond to higher temperatures.
For lower metallicities and at the extreme of lower masses
($M\la0.7$~\Msun), stars also develop higher temperatures, going
rapidly to the region of the blue HBs (eventually closing the ZAHB
line in the HR diagram, and crossing the RR Lyrae instability strip;
see e.g.\ the behaviour of $Z=0.001$ models in \reffig{fig_zahb}).
The lower panel of \reffig{fig_zahb} shows the theoretical results
transformed into the observed $M_I$-$(V-I)_0$ plane (see Section~3.1).

It is convenient to limit our discussion to only those stars which can
be relevant in defining the red clump morphology.  Only $Z\ga0.004$
models, for instance, can produce red clumps in the \vi\ colour range
under discussion (see \reffig{fig_zahb}). 
Moreover, $Z\ga0.004$ stars with ZAHB masses lower than
about 0.8~\Msun\ are hardly present in the stellar populations we will
consider, since they (i) neither live long enough for reaching the
He-burning phase during a Hubble time, (ii) nor are produced by mass
loss on the RGB of more massive stars, if we assume that Reimers'
(1975) mass loss prescription is valid. Therefore we can further limit
the analysis to stars with $M\ga0.8$~\Msun.  What turns out is that,
for a single value of $Z$, clump stars of relatively high masses (and
hence younger) are systematically hotter than those of lower masses
(and hence older). Moreover, the bluest are also the faintest among
the clump stars.

\subsection{Mass distribution of clump stars}

When we deal with composite stellar populations, it is also important
to consider what is the relative probability of populating each
section of the clump. This can be estimated from the following:

First, if we neglect mass loss on the RGB, we have that at any age
clump stars have masses \Mhe\ similar to turn-off ones, i.e.\
$\Mto=\Mhe$. The probability of having ZAHB masses in the interval
$[\Mhe,\Mhe+\diff\Mhe]$ is then proportional to the product of the
initial mass function (IMF) $\phi_{M}$ with the star formation rate at
the time the stars formed, $\psi(T-t)$, where $T$ is the galactic age.
For our $Z=0.019$ models, TO-ages are related to masses according to
\beq 
\logt\simeq10.14 - 3.25\log(\Mto/\Msun) \,. 
\label{eq_life} 
\eeq 

Core He-burning stars evolve from the ZAHB to higher luminosities
leaving the clump after a time interval  $t\sub{He}$.  They grow in
luminosity by only about $0.05-0.1$~mag, before the faster evolution
towards the early-AGB starts. Therefore, the resulting probability of
detecting a star of mass \Mhe\ in the clump is proportional to
$\phi_{M}(\Mhe)\,\psi[T-t(\Mhe)]\,t\sub{He}(\Mhe)$.  The function
$t\sub{He}$ assumes almost constant values, close to $10^8$~yr, for
the entire range $\Mhe\la\Mhef$, but nearly doubles for masses slightly
above this limit (Girardi \& Bertelli 1998).

Thus, in the case of constant star formation, the mass distribution of
clump stars closely follows the IMF, except for those stars with mass
slightly higher than \Mhef, the detection of which would be favoured by a
factor of two with respect to this rule.  For instance, if we assume
$\psi(t)$ constant in the solar neighbourhood from 1 to 10 Gyr
(corresponding to \Mhe\ from 2.2 to about 1~\Msun), the local sample of
stars would contain only 2.5 times less clump stars with mass in the
interval $2.1>(\Mhe/\Msun)>2.0$ than in the $1.1>(\Mhe/\Msun)>1.0$
one.

Mass loss on the RGB causes stars of a given age to have
lower values of \Mhe\ than given by \refeq{eq_life} under the
assumption that $\Mto=\Mhe$. However, the effect is not dramatic. If
we adopt Reimers' (1975) mass loss rates with an efficiency parameter
$\eta=0.4$ (Renzini \& Fusi Pecci 1988, and references therein), we
conclude that only the long-lived stars with initial masses
$M\la1.2$~\Msun\ have their masses reduced by more than
0.1~\Msun. Stars of masses higher than about 1.5~\Msun\ are virtually
unaffected.  Therefore, mass loss causes a spread of the stars in the
red extremity of the clump to slightly lower masses and hence
luminosities (see \reffig{fig_zahb}), while the bluest part of the
clump remains essentially the same.

\section{Clump distribution in the \mi\ vs.\ \vi\ plane}
\label{sec_clump}

\subsection{Theoretical predictions}

All effects described above, which determine the observed
distribution of clump stars in the HR diagram of composite stellar
populations, can be  simulated by means of evolutionary
population synthesis models. The code we have for this purpose (see
Girardi \& Bertelli 1998) is based on the construction of isochrones
for different ages and metallicities (interpolations in both
parameters are allowed), which are then populated with stars and
summed up in order to produce the theoretical colour-magnitude
diagrams (CMD) and luminosity functions (LF). Mass loss along the RGB
is taken into account during the stage of isochrone construction, and
given by the Reimers' (1975) formula multiplied by a factor
$\eta=0.4$. Every point in the isochrones is transformed to the
observational colours and magnitudes according to the Kurucz (1992)
transformations. The latter produce some systematic shifts in the
transformed colours, which have so far been well documented (e.g.\
Worthey 1994; Gratton, Carretta \& Castelli 1996; Charlot, Worthey \&
Bressan 1996). Fortunately, the inadequacies in Kurucz transformations
are not critical in the \vi\ colour, being smaller than 0.1~mag for
both dwarfs and giants of solar metallicity (Gratton et al.\ 1996).

\begin{figure}
\begin{center}
\psfig{file=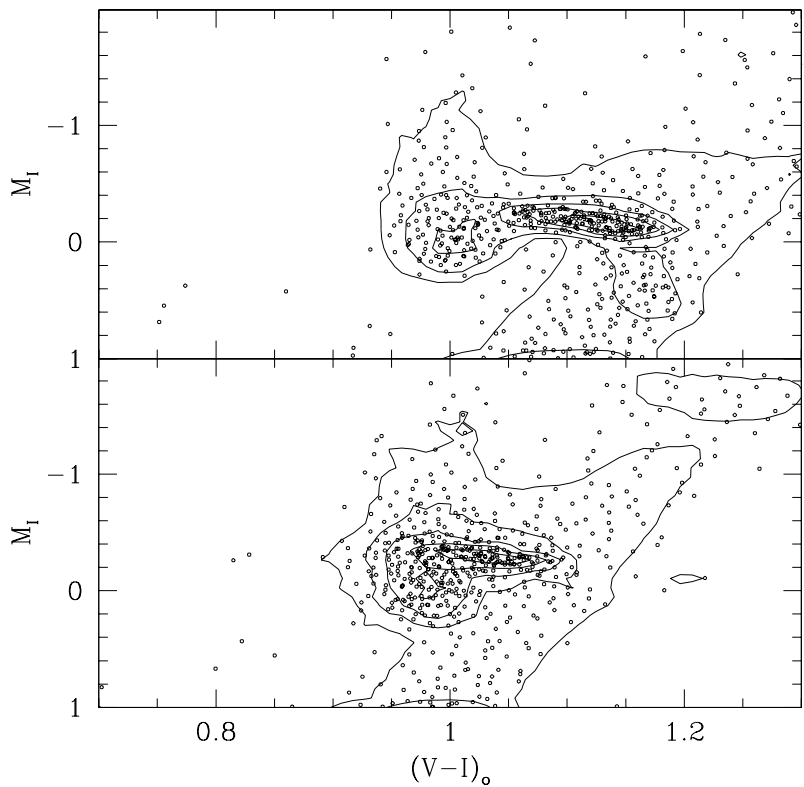,width=11cm}
        \caption{Theoretical distribution of clump stars in the \mi\
vs.\ \vi\ diagram. We simulate a total of 600 stars (dots) following
the predicted density distributions.  5 countour levels (continuous
lines) delimit regions with the same density of stars.  This model is for a
composite stellar population with constant star formation rate in the
interval $0.1<(t/{\rm Gyr})<10$ and mean solar metallicity (upper
panel), and for another one with constant star formation but assuming
an age metallicity relation and larger scale heights for older stars
(see text for details). }
\label{fig_vvi}
\end{center}
\end{figure}

\reffig{fig_vvi} shows the result of the simulations in the \mi\ vs.\
\vi\ diagram.  Illustrated in the top panel is the result of summing
up stellar populations with ages from 0.1 to 10
Gyr (younger stellar populations certainly do not contribute to the
clump), with mean solar metallicity ($Z=0.019$) and a metallicity
dispersion of only $\sigma(\log Z)=0.1$. 
It is meant to represent, as a first approximation, the bulk
of local stars sampled by {\em Hipparcos}. No attempt was made to
simulate dispersions in reddening and distance, observational errors,
and the presence of multiple stars. Any age-metallicity relation was
also ignored in this first simulation. In order to obtain more
realistic models, we make a second simulation, this time assuming that
the metallicity decreases with age at a rate of $-0.07$ dex/Gyr
(Carraro, Ng \& Portinari 1998), and correcting for the relative number of
stars with different scale heights above the galactic plane, according
to the $f(Z)$ function indicated by the $[\epsilon=0.02, P=1.8]$ case in
Sommer-Larsen (1991). The result is presented in the bottom panel of
\reffig{fig_vvi}.

In both cases, the theoretical clump presents a main almost-horizontal
feature, ranging from $\vi=1.0$ to $\vi=1.1-1.2$, and a secondary,
predominantly vertical feature, developing at the extreme left of the
clump. This secondary structure is caused by the stars with masses
above $\Mhef\simeq2$~\Msun, and its maximum density occurs
$\sim0.3$~mag {\em below} the main body of the clump. Also, it originates a
kind of plume departing from the clump to higher luminosities.
The horizontal part appears to be a generic feature of the local
clump. 

\begin{figure}
\begin{center}
\psfig{file=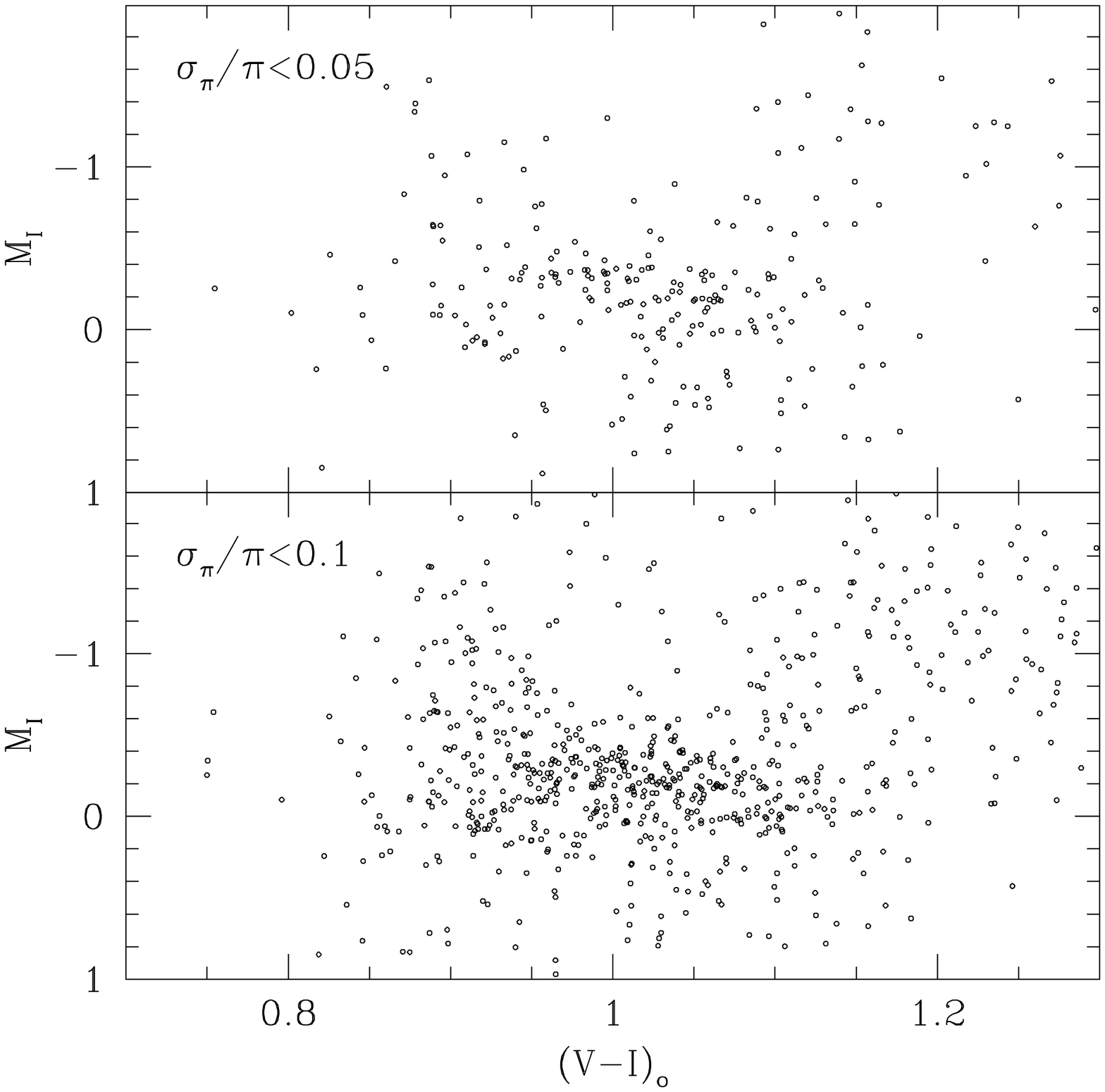,width=11cm}
        \caption{Distribution of clump stars in the \mi\ vs.\ \vi\
diagram, from {\em Hipparcos} data. The sample is limited to data with
parallax errors smaller than 5\% (upper panel) and 10\% (lower one).
In order to suppress artificial superposition of points in the diagram
(due to limited digits in the database), random numbers between
$-0.005$ and 0.005 were added to the \vi\ colour.}
\label{fig_hippvi}
\end{center}
\end{figure} 

\subsection{The clump from {\em Hipparcos} data}
\label{sec_hippa}

From the {\em Hipparcos} database we have selected stars that
fulfilled the following criteria: error in parallax $<10$\%, a
parallax $>0$, the `data-points-rejected' (DPR, field H29) flag
$<10$\%, `goodness-of-fit' (GOF, field H30) flag $<3$, the `number of
components' (field H58 in the database) equal to 1, and the `source
for $I$ photometry' flag (field H42) either A, C, E, F or G. This
results in 2546 stars. These are the same selection criteria as
Paczy\'nski \& Stanek (1998), except that we added the criteria on
DPR, GOF and field H58 to eliminate bad solutions and to a large
extent binary stars.  The red clump derived from these data is plotted
in \reffig{fig_hippvi}, and limited to samples with parallax errors
smaller than 5\% (upper panel) and 10\% (lower panel). Therefore,
absolute magnitudes for these two samples are known to better than
0.11 and 0.22~mag (standard errors), respectively. The \vi\ colours
are also expected to be accurate to within 0.05~mag, since reddening
is negligible for them; the mean error in \vi\ for this dataset is of
order 0.02~mag.

Several aspects of the observed red clump in \reffig{fig_hippvi} find
some correspondence with the structure of the theoretical one, shown in
\reffig{fig_vvi}. In particular, it is clear that the observed stars
are more spread in \mi\ at the blue side of the clump than at the
red. Also, there is a locally higher concentration of stars at
$[\mi\simeq0.0, \vi\simeq0.92]$, which is more evident in the plot
which considers data with parallax errors smaller than 5\%.  Also, a
plume of brighter stars is present at $\vi\simeq0.9$, going from
$\mi\simeq-0.5$ to about $-1.0$. These latter features nicely
correspond to those predicted to occur at the blue side of the clump.

However, it is also evident that the theoretical distribution is
systematically shifted to redder colours with respect to the observed
one. In order to make both coincide, a shift of about
$\Delta(\vi)=-0.08$ to the models is required; that is of the same
order of the inadequacies found to date in the \vi\ colour
transformations of Kurucz (1992).
 
The plume departing from the clump to higher luminosities has been
already noticed by Beaulieu \& Sackett (1998) in $BV$ data from the
{\em Hipparcos} database, 
and an equivalent feature in the CMD of the LMC has been at
the center of a recent debate about its origin (Zaritsky \& Lin 1997;
Beaulieu \& Sackett 1998). The structure located
slightly below the main red clump, at its bluest extremity,
passed however unnoticed even in works which dealed with the
clump morphology from {\em Hipparcos} data (e.g.\ Jimenez et al.\ 1998).
Before going ahead, it is important to be sure of the reality of this
kind of `secondary clump'.

\begin{figure}
\begin{center}
\psfig{file=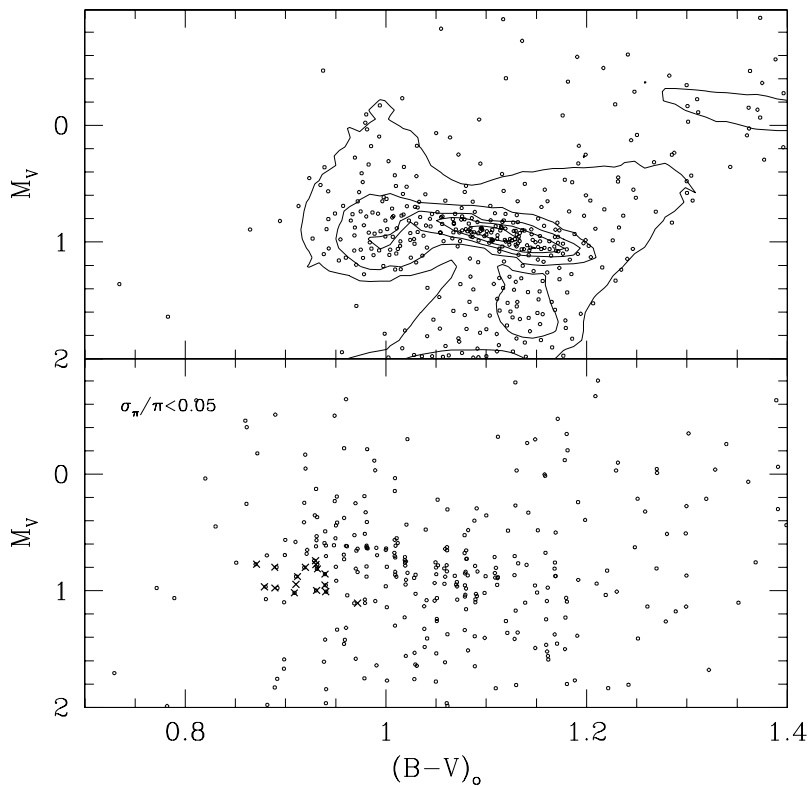,width=11cm}
        \caption{Distribution of clump stars in the \mv\ vs.\ \bv\
diagram, from theoretical models (upper panel; corresponding to the
model illustrated in the upper panel of \reffig{fig_vvi}, for a total
number of 400 stars), and
{\em Hipparcos} data with parallax errors smaller than 5\% (lower
panel). Stars belonging to the secondary clump in the \mi\ vs.\ \vi\
diagram (with $0.84<(\vi)<0.94$ and $0.2>\mi>-0.2$, 
see \reffig{fig_hippvi}) are indicated by crosses.}
\label{fig_hippbv}
\end{center}
\end{figure} 

In \reffig{fig_hippbv}, we show how the red clump shows up in the \mv\
vs.\ \bv\ diagram, as predicted by the model which does not consider
the age-metallicity relation (upper panel), and as observed by {\em
Hipparcos} (lower panel, sample limited to 5\% parallax errors). 
Notice that also in this plot models are
systematically redder by about $\Delta(\bv)=0.08$ with respect to 
observations. In
the lower diagram, we identify by crosses the stars which were located in the
small box with $0.84<(\vi)<0.94$ and $0.2>\mi>-0.2$ in the previous
\reffig{fig_hippvi}. It can be noticed that the red clump is expected
to be less resolved in this diagram. In addition it is
inclined, due to the stronger dependency of bolometric corrections on
the \bv\ colour.  Nonetheless, 
the stars that define the secondary clump in \reffig{fig_hippvi} again
occupy a well-defined region of the \mv\ vs.\ \bv\ diagram, being
located to the blue and to a higher magnitude (i.e. are fainter) with
respect to the main 
body of the red clump at that colour.  The statistics provided in this
plot is higher than that derived from the equivalent one in
\reffig{fig_hippvi}. 

We conclude that the predition of the theoretical models for the
secondary clump appears to be realistic, since the substructure is
observed in both CMD diagrams, and it is unlikely that it might be
caused by such effects as reddening, bad number
statistics, contamination from the RGB and low-metallicity clump stars, 
photometric and parallax errors.

\newpage
There are two remarkable aspects in this feature:
\begin{enumerate}
\item It represents an old prediction of stellar models, that the core
masses at helium ignition has a minimum value for $M\ga\Mhef$,
corresponding to a minimum value of luminosity.  As far as we are
aware, the only observational search for this feature was carried out
by Corsi et al.\ (1994), who successfully located a minimum in the
$V$ luminosity of clump stars for LMC clusters about 0.6 Gyr old
(corresponding to 1~Gyr if moderate overshoot is assumed). In Girardi
\& Bertelli (1998), the present models with $Z=0.008$ are shown to
provide a good general description of Corsi et al.\ (1994) data in
regard to this point. The presence of a similar feature in {\em
Hipparcos} data provides further confirmation of theoretical
predictions.
\item The observation of such a feature in a local, nearby sample of
stars opens the possibility of further testing stellar models. As
discussed in \refsec{sec_stars}, stars in this secondary clump
correspond to a quite limited interval of stellar masses, which goes
from \Mhef\ to about $\Mhef+0.3$~\Msun. Therefore, the direct
measurement of their masses could provide stringent constraints to
\Mhef, and therefore on the efficiency of convective core overshoot
for stars of similar masses.
\end{enumerate}

In order to check the latter point, we looked for binary stars in the {\em
Hipparcos} catalog, in which the primary could be considered a member
of this secondary clump.  We found 7 binaries  contained in the
$0.84<(\vi)<0.94$ and 
$0.2>\mi>-0.2$ box, two of which are visual ones with reliable orbital
parameters in the literature:
\begin{itemize}
\item
For HD~41116, the orbital parameters by Heintz (1986) and 
{\em Hipparcos} parallax
provide a total mass of $4.30\pm0.66$~\Msun\ for the system. As
both primary and secondary are evolved (K0III+G8III,IV:
according to Stephenson \& Sanwal 1969), their masses would be similar
and of about 2.1~\Msun\ each. A complication arises from the
later-type star being itself a spectroscopic binary (WDS catalog;
Worley \& Douglass 1997).  
\item
HD~90537 is a K0III,IV+F8V binary (Edwards
1976), with a magnitude difference of 1.42~mag between both components
({\em Hipparcos} catalog).  The orbital parameters and fractional mass
given by Heintz (1982), together with {\em Hipparcos} parallax,
indicate a mass of $1.92\pm0.34$~\Msun\ for the primary, and
$0.53\pm0.16$~\Msun\ for the secondary. The mass of the primary is
consistent with the interpretation that it should have a mass close to
\Mhef. However, the low value of mass obtained for the secondary seems
in contradiction with its estimated spectral type, F8V. 
\end{itemize}
The above numbers do not give a clear indication whether the
primary stars should belong to the secondary clump we found, and
whether their masses are really close to 2~\Msun. In both
cases, the precise classification of the component stars would be
necessary in order to better constrain the characteristics of the
primary. A systematic study of them would be of
high interest.

\section{Distance determinations by means of the red clump}
\label{sec_lf}

In most of the works that used the red clump as a standard candle, the
LF of clump stars has been modeled by fitting the function
\beqa
N(\mi) & = & a + b\mi + c\mi^2 + \nonumber \\
	& &\frac{N_{\rm tot}}{\sigma\sqrt{2\pi}}
	\exp\left[-\frac{(\mi-\mimax)^2}{2\sigma^2}\right] \; .
\label{eq_gauss}
\eeqa
over a limited \mi\ and \vi\ interval. The fit to {\em Hipparcos} data provides
the reference value of \mimax. Similarly, a value of \imax\ can be
derived for the red clump in distant stellar systems, and the distance
modulus $\dm=\imax-\mimax$ is obtained in a single step.

\subsection{Calibrating \mimax\ from {\em Hipparcos} data}

Following Paczy\'nski \& Stanek (1998), we have modelled the {\em
Hipparcos} data by a chi-square fit of \refeq{eq_gauss}. In order to
generate solutions with acceptable 
chi-square residuals, fittings were limited to the $1.0>\mi>-1.5$
interval.

\begin{figure}
\centerline{
\psfig{file=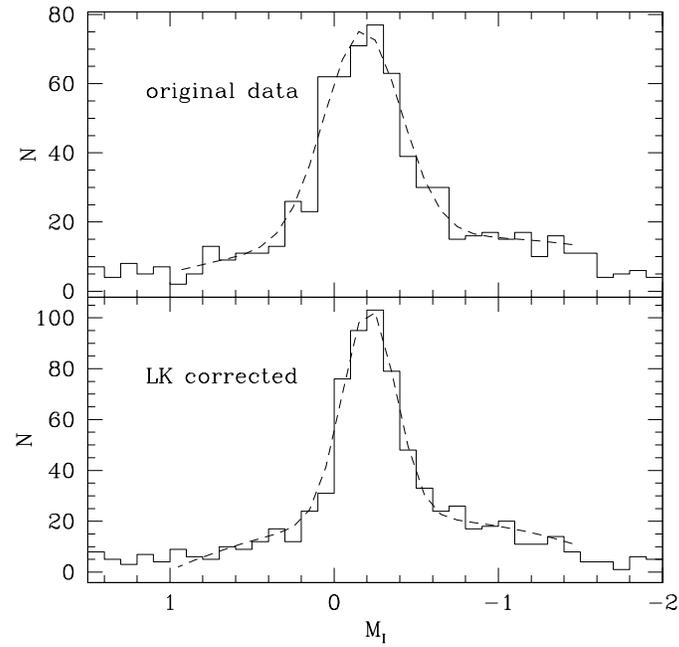,width=9cm}}
        \caption{LFs for clump stars in {\em Hipparcos} data with parallax
errors smaller than 10\% (continuous line: histogram of the data;
dashed line: chi-square fit of equation \ref{eq_gauss}). The upper panel
presents the original data in the $0.8<(\vi)<1.25$ interval, while the
lower one presents the data corrected by Lutz-Kelker bias, as
discussed in the text. }
\label{fig_lfhipp}
\end{figure}

\reffig{fig_lfhipp} presents the results for the data in the
$0.8<(\vi)<1.25$ colour interval, sampled with a 0.1~mag resolution
(upper panel).  We obtain for this distribution a best fit with
$\mimax=-0.177$ and $\sigma=0.226$.  These values agree, within the
errors, with those obtained by Paczy\'nski \& Stanek (1998) and
Stanek \& Garnavich (1998), i.e.\ $\mimax=-0.185\pm0.016$ and
$\sigma=0.243$, from similar data. The small differences in the
numbers probably reflect the different approaches in the fitting
procedures, and the small differences in the selected samples.

It is worth remarking that the sample of red clump stars here used is
essentially a complete one. Indeed, stars in the clump and with
parallax errors smaller than 10\% are always at least 0.5~mag brighter
than the $V\simeq8.5$ completeness limit of the {\em Hipparcos}
catalogue. 

The bottom panel in \reffig{fig_lfhipp} presents the LF for the {\em
Hipparcos} data when corrected by the Lutz-Kelker bias according to
Salaris \& Groenewegen (1998). This procedure follows Turon Lacarrieu
\& Cr\'ez\'e (1977) and Smith (1987) in that it assumes one has
a-priori knowledge about the absolute magnitude of the sample of
stars, described by a Gaussian distribution with a mean and error. The
Lutz-Kelker correction is a statistical one and therefore one can also
calculate the error in the correction. This novelty is outlined in
Salaris \& Groenewegen (1998).

First, main-sequence stars and giants were separated. Either by the
luminosity classification and the spectral type 
listed in the {\em Hipparcos} database, or
on colour and absolute magnitude using the {\em Hipparcos} parallax as
listed. A star was considered to be a dwarf if $\mi >
-2.0+5.333\,(\vi)$, or if $\mi>5$ and $\vi>1.5$.  For all stars
considered to be dwarfs we used as theoretical input value in the
Lutz-Kelker correction an empirical value for \mi\ as a funtion of
\vi\ colour, based on the {\em Hipparcos} data for stars with a
parallax better than 5\%. The assumed spread in the theoretical value
was estimated by eye based on the spread in the data, and is 0.35~mag
for $\vi<0.75$, and 0.15~mag else. All other stars are considered to
be clump giants and a mean and spread of $-0.18$ and 0.24 are used.

Thus, after correcting the \mi\ data in this way, we obtain a slightly
brighter, and significantly sharper clump, with $\mimax=-0.209$ and
$\sigma=0.160$ (\reffig{fig_lfhipp}).

\begin{figure}
\centerline{
\psfig{file=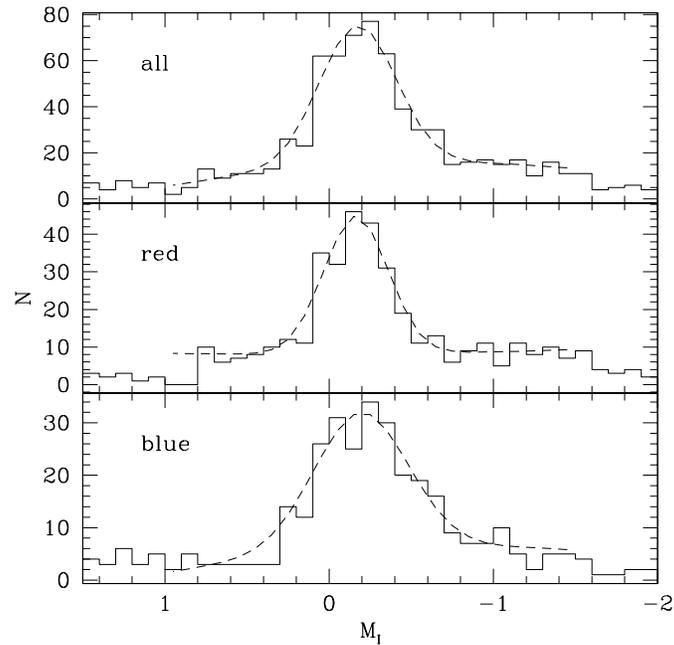,width=9cm}}
        \caption{LFs for clump stars in {\em Hipparcos} data, as in
\reffig{fig_lfhipp}. We present the data in the complete
$0.8<(\vi)<1.25$ interval (upper panel); and divided into red
($1.0<(\vi)<1.25$; middle) and blue ($0.8<(\vi)<1.0$; lower) samples. }
\label{fig_lfhipp_br}
\end{figure} 

\reffig{fig_lfhipp_br} instead presents LFs for the original data
(without LK correction) in
the $0.8<(\vi)<1.25$ interval (upper panel), and sampled in the two
separate $1.0<(\vi)<1.25$ (middle panel) and $0.8<(\vi)<1.0$ (lower
one) colour intervals, at 0.1~mag resolution. For the two cases we find
$\mimax=-0.173$ and $\sigma=0.201$, respectively $\mimax=-0.196$ and
$\sigma=0.289$.  What is remarkable in these histograms
is the higher dispersion $\sigma$ in the blue part of the clump.  This
high dispersion is interpreted as the result of evolutionary effects,
as discussed below. 

\subsection{Model predictions for \mimax}
\label{sec_model_mimax}

We are now going to discuss what, according to
our models, is the expected LF of clump stars.
The aim is to cast light on the factors which give origin
to the observed LFs shown in Figs.~\ref{fig_lfhipp} and
\ref{fig_lfhipp_br}, and which can also influence the derived values
of \mimax. We start the comparison with the simple model depicted in
the upper panel of \reffig{fig_vvi}, which gives a reasonable
description of the clump in the \mi\ vs.\ \vi\ diagram. However, our
models are first corrected by the colour shift $\Delta(\vi)=-0.08$,
otherwise the comparison with observed \mimax\ values would be
meaningless. 

\begin{figure}
\centerline{
\psfig{file=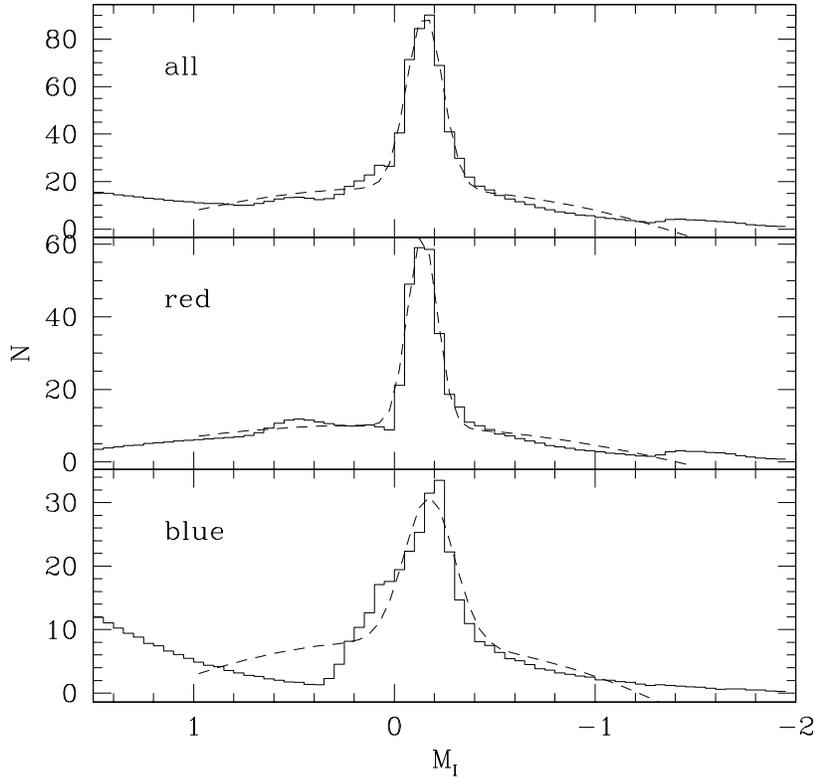,width=11cm}}
        \caption[]{Theoretical LFs for the simulation depicted in the
upper panel of \reffig{fig_vvi}. The original simulation was,
previously, shifted by $\Delta(\vi)=-0.08$, in order to reproduce the
colour range appropriate to the red clump observed by {\em
Hipparcos}, and the relative numbers of red and blue stars. 
Similarly to \reffig{fig_lfhipp_br}, we present the LFs
for 3 different colour intervals: for the complete $0.8<(\vi)<1.25$
one (upper panel); and divided into red ($1.0<(\vi)<1.25$; middle) and
blue ($0.8<(\vi)<1.0$; lower) samples. The LFs are normalized to the
observed distributions (\reffig{fig_lfhipp_br}).}
\label{fig_simbr}
\end{figure} 

In \reffig{fig_simbr}, we present the synthetic \mi\ LF for different
cases, representing all stars, and those in the red or blue parts of
the clump. Again, `blue' and `red' parts are defined by stars in the
colour intervals $0.8<(\vi)<1.0$ and $1.0<(\vi)<1.25$,
respectively.  We checked that the shift of $\Delta(\vi)=-0.08$
applied to the models allows us to reproduce
the same relative number of blue and red clump stars as in the {\em
Hipparcos} sample of \reffig{fig_lfhipp_br}.

The fit of \refeq{eq_gauss}
describes well the central spike due to clump stars, at $0>\mi>-0.4$,
but not the wings. Both wings are essentially due to evolutionary
effects: the bright one, extending up to about $\mi\simeq-0.9$, is
caused by stars which have evolved already from the ZAHB, while
the faint one is caused mainly by the stars with $M\ga\Mhef$,
located at about $0.3>\mi>-0.1$, and appears only in the blue colour bin. In
the red bin, there is another, fainter bump-like feature below the
clump, at $0>\mi>0.6$, which corresponds to the bump of the RGB stars
(see e.g.\ Renzini \& Fusi Pecci 1988; Cassisi \& Salaris
1997). Another interesting feature present in the plot is the bump in
the LF caused by the presence of early-AGB stars, located at
$-1.3>\mi>-1.8$, in the red bin. The latter two features are
conspicuous in CMD studies of galactic globular clusters (e.g.\ Fusi
Pecci et al.\ 1990; Buonanno, Corsi \& Fusi Pecci 1985).

For this theoretical simulation, we obtain $\mimax=-0.151$,
$\sigma=0.089$ for all stars, $\mimax=-0.140$, $\sigma=0.073$ for red
ones, and $\mimax=-0.174$, $\sigma=0.123$ for blue
ones. It is clear
that the $\sigma$ values obtained are about 1/3 of the observed ones
(see \reffig{fig_lfhipp_br}). Also, blue stars have a \mimax\ value
$-0.034$~mag brighter than red ones.

The simulations cannot be simply compared to observations, since
the latter include the errors in parallax and hence in distance
modulus. Therefore, we have convolved the theoretical distribution of
\reffig{fig_simbr} with the expected distribution of errors in
distance modulus resulting from the errors in {\em Hipparcos}
parallaxes (see Appendix).

\begin{figure}
\centerline{
\psfig{file=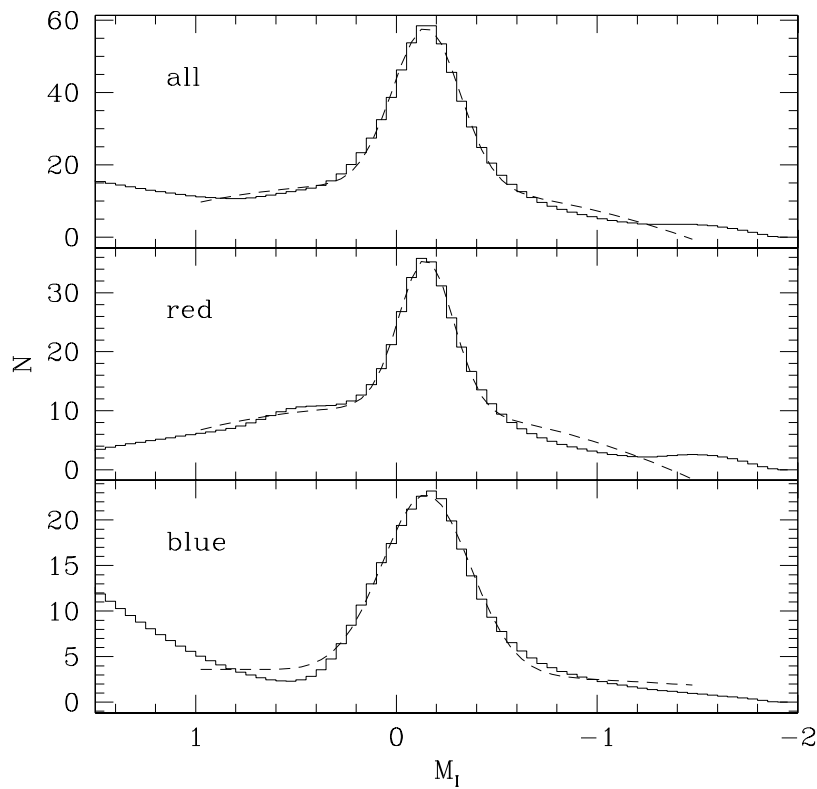,width=11cm}}
        \caption{LFs for the simulations depicted in
\reffig{fig_simbr}, this time convolved with the function
$f(\deltadm)$ (\refeq{eq_conv}) shown in \reffig{fig_ddm}. These LFs
should be compared to the observed ones, shown in
\reffig{fig_lfhipp_br}.}
\label{fig_simconv}
\end{figure} 

The results of this procedure are shown in \reffig{fig_simconv}.  This
figure 
demonstrates that the derived LFs can be fitted much better to the
\refeq{eq_gauss} than previously. We get $\mimax=-0.149$,
$\sigma=0.167$ for all clump stars, $\mimax=-0.148$, $\sigma=0.137$
for the red ones, and $\mimax=-0.149$, $\sigma=0.22$ for the blue
ones. Clearly, the $\sigma$-values are close to the observed ones
obtained in the previous section (and \reffig{fig_lfhipp_br}).  
$\mimax$ for the whole sample and the red colour bin are not influenced
by the convolution. However,
particularly interesting is that the
faint wing of the LF for the blue sample (at $0.3>\mi>-0.1$ in 
the lower panel of \reffig{fig_simbr}), after the
convolution, becomes indistinguishable from the main body of the
clump. In this case, the fitting routine includes those fainter stars
into the gaussian curve. Therefore, a
fainter \mimax\ is obtained for the blue stars.  The \mimax\
difference between blue and red stars becomes null
due to this effect.

Moreover, this theoretical red clump LF shares several characteristics
with the observed one: very nice gaussian profiles for the
central clump; the presence of extended wings which end at about
$\mi=0.8$ at the faint side, and at $\mi=-1.8$ in the bright one;
comparable values of $\sigma$ and \mimax; a larger $\sigma$ for the blue
stars; and, most important, \mimax\ values almost independent of \vi.
This latter characteristic is essentially due to the fact that the
stars at the bluest extreme of the red clump are of higher masses,
thus being on the mean younger and fainter than the remaining clump
stars. As they are, in general, not clearly separated in \mi\ from the
main red clump (especially for the sample with $<10$\% parallax
errors), they have a weight in determining the mean clump magnitude
\mimax.  How much \mimax\ can be affected by the young stars, however,
depends essentially on two factors:

\begin{enumerate}
\item On the relative proportion of young (ages around 1~Gyr) to old
($>1$~Gyr) stars. Comparison of Figs.~\ref{fig_simconv} and
\ref{fig_lfhipp_br} reveals that the simulation with constant star
formation rate provides almost the same relative number of stars in
the red and blue parts of the clump as in the {\em Hipparcos} sample. This
could suggest that the star formation rate 1~Gyr ago was the same as
the mean over the Galactic Disk history. Of course, any conclusion of
this kind requires closer scrutiny by means of more realistic models
for the age-metallicity relation, initial mass function, and scale
heights of different stars, which is beyond the scope of this paper.
\item Whether these stars fall inside or outside the colour interval
defined to measure \mimax.  In the case of {\em Hipparcos} data, the
complete red clump falls in the $0.8<(\vi)<1.25$ interval used to define
the reference \mimax\ by Paczi\'nski \& Stanek (1988). In the case of
stellar populations with a lower mean metallicity as the Magellanic
Clouds, the red clump is observed at much bluer colours, and falls
partially out of the $0.8<(\vi)<1.25$ interval.  For instance, 
Fig.~2 in Stanek et al.\ (1998) shows that the red clump in the LMC
ranges from $\vi=0.6$ to 0.9. In this case, only the reddest tail of
the clump falls inside $0.8<(\vi)<1.25$, which probably excludes the
younger (and fainter) stars from their estimate of \imax\ for this
galaxy.
\end{enumerate}

Both factors mean that there can be systematic effects in the \imax\
measurement, that are determined by the way we sample the clump stars
and by the relative contribution of young stars to the composite
stellar population observed.

Our models reproduce the fact that \mimax\ is almost constant over
the \vi\ interval considered
in the case of {\em Hipparcos} data, and yet predict a significant
dependence of the clump magnitudes with metallicity
(\reffig{fig_zahb}). This is mainly due to the dependence of the
underlying stellar models' luminosity on metallicity. For the
transformation into \mi, we only have to assume that the bolometric
corrections given by Kurucz (1992) have the correct (weak)
dependence on metallicity.

\section{The expected \mimax\ for the LMC}
\label{sec_lmc}

The history of star formation in the LMC has been intensively studied
in the last years (Bertelli et al.\ 1992; Vallenari et al.\ 1996;
Gallagher et al.\ 1996; Holtzman et al.\ 1997; Elson, Gilmore \&
Santiago 1997; Stappers et al.\ 1997), by means of CMDs of different
fields. Most of these works point out that the star formation rate in the
LMC was higher in the last few Gyr than in the interval from about 3
to $\sim12$~Gyr. That is in agreement with the fact that no cluster is
known with ages in the $4-10$~Gyr interval (van den Bergh 1981; Da
Costa 1991; Girardi et al.\ 1995; Olzsewski, Suntzeff \& Mateo 1996;
Geisler et al.\ 1997). Elson et al.\ (1997) present additional
evidences that a later burst of stellar formation has taken place
about 1~Gyr ago, while Geha et al.\ (1998) propose that a continuous
stellar formation history was followed by an enhancement of star
formation from 2~Gyr to now.

With respect to the abundances of LMC stars, the results from CMD
studies are not conclusive in giving a clear age-metallicity relation
for the LMC. This applies also to clusters: while Olszewski et al.\
(1991) pointed out that intermediate-age LMC clusters have a mean
$\feh=-0.4$ (or $Z=0.008$), there are several claims in the literature
that a significantly lower value would be more appropriate (see
Bica et al.\ 1998, and references therein).

\begin{figure}
\begin{center}
\psfig{file=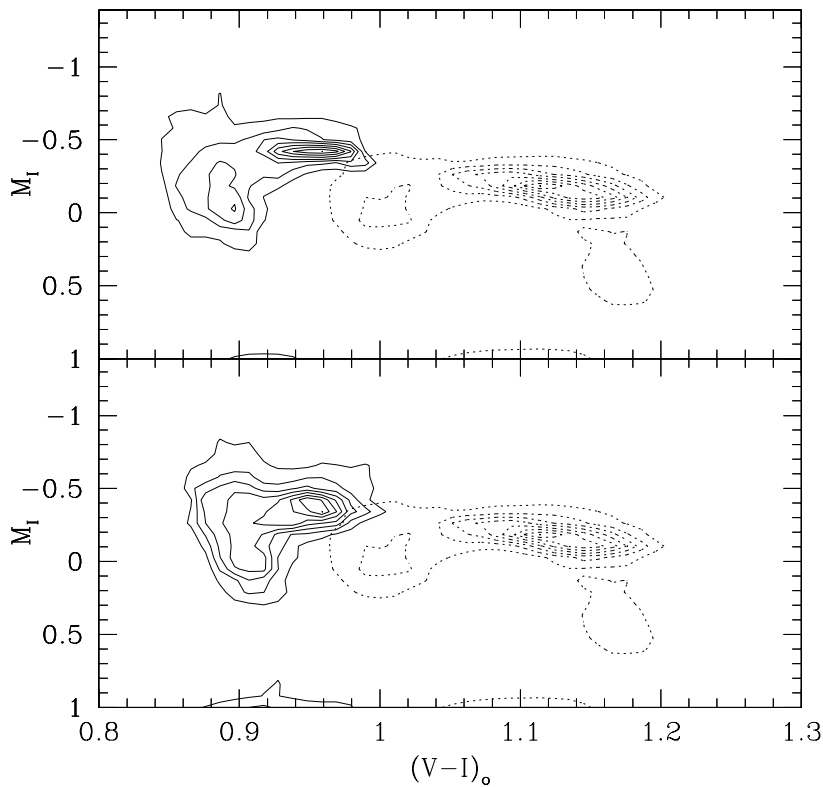,width=11cm}
        \caption{Theoretical density of stars in the \mi\ vs.\ \vi\
diagram, for the models meant to represent the LMC data (continuous
lines). For the sake of comparison, the model for the {\em Hipparcos}
data (\reffig{fig_vvi}) is also repeated (dotted lines). Only the
contour levels corresponding to the red clump are shown. Upper
panel: model with $[0.004<Z<0.008, 0.1<(t/{\rm Gyr})<3]$; lower panel:
model with a star formation and chemical enrichment history similar to
that suggested by Vallenari et al.\ (1997; see text for details).  }
\label{fig_vvi_lmc}
\end{center}

\end{figure} 

Given the variety of solutions suggested by several authors, we prefer
to limit our analysis to two scenarios. For them,
\reffig{fig_vvi_lmc} shows the derived stellar density in the \mi\
vs.\ \vi\ plane, compared to that meant to represent {\em Hipparcos}
data. The first one assumes simply a constant star formation rate
between ages 0.1 and 3~Gyr, with metallicities equally probable
between $Z=0.004$ and 0.008.  This model provides a clump with some
structure, especially in its blue part, and located at the border of
the $0.8<(\vi)<1.25$ interval (again, if a $\Delta(\vi)=-0.07$ shift is
previously applied to the models).  Following the suggestion by Cole
(1998), we tested a second scenario  similar to that of Vallenari et
al.\ (1996): 44\% of the clump being produced by $[0.006<Z<0.010,\,\,
0.1<(t/{\rm Gyr})<2]$ populations, the remaining 56\% by
$[0.002<Z<0.006,\,\, 2<(t/{\rm Gyr})<10]$ ones. Indeed, this alternative
provides a model more similar to the apparently featureless clump
observed by Stanek et al.\ (1998). The scenario proposed by Holtzman
et al.\ (1997), instead, 
would produce a clump separated into a blue [with $Z=0.001$ and
$2<(t/{\rm Gyr})<10$] and a red [with $Z=0.008$ and $1<(t/{\rm
Gyr})<2$] component, contrary to observations; therefore we
preferred not to test this possibility here.

As simple as these simulations are, they indicate a clump located
about 0.2 mag to the blue in \vi, and about 0.2~mag brighter in \mi\
than the local one sampled by {\em Hipparcos}.  Comparison of this
clump with the observed one presented by Stanek et al.\ (1988) is not
conclusive. In fact, their data show a featureless clump, about
0.3~mag wide in \vi\ and with a $\sigma(I)$ of about 0.15~mag. Part of
this width can be attributed to the dispersion in reddening in the
observed fields. The standard errors in the determination of the mean
$E_{V-I}$ and $A_I$ by Stanek et al.\ (1998), $\sigma(\evi)=0.05$ and
$\sigma(A_I)=0.08$, represent a lower limit to the magnitude of this
dispersion.  Models in \reffig{fig_vvi_lmc} do not include these
effects.  Moreover, we remark that attempts to derive the best clump
model should better take into account the distribution of stars in
the complete CMD, and not only in the clump, in order to constrain the
possible solutions. Again, this is beyond the scope of this paper.

\begin{figure}
\begin{center}
\psfig{file=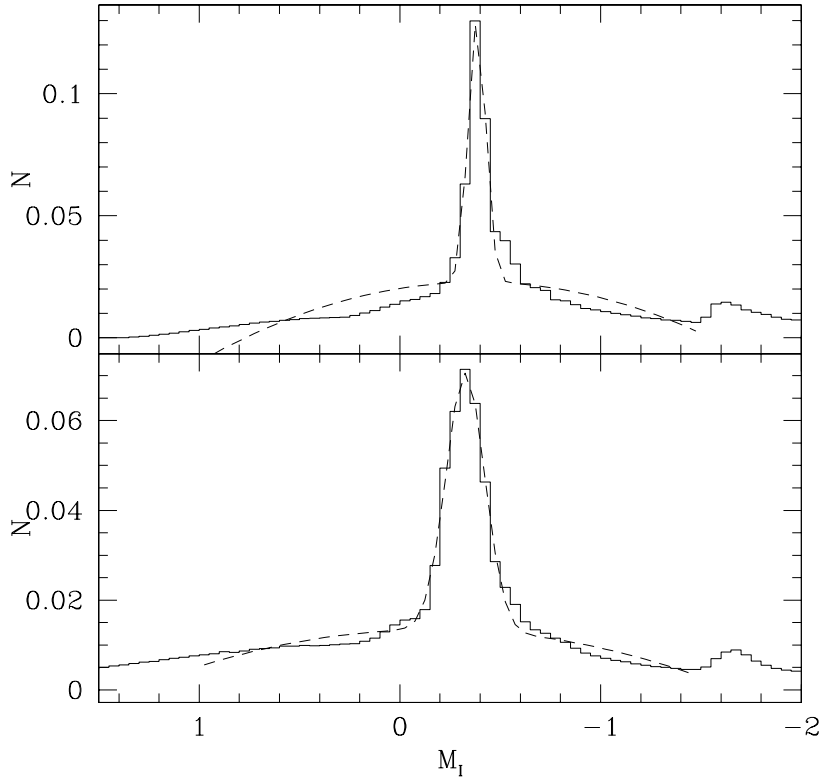,width=11cm}
        \caption{Theoretical LFs for the LMC models (solid lines),
together with the fit of \refeq{eq_gauss} (dashed lines). Upper and
        lower panels correspond to those of \reffig{fig_vvi_lmc}. }
\label{fig_lf_lmc}
\end{center}
\end{figure} 

The LF for these theoretically predicted clumps are presented in
\reffig{fig_lf_lmc}. Fits to these curves according to
\refeq{eq_gauss} produce values of
$\mimax=-0.384$, $\sigma=0.043$ for the simple simulation, and
$\mimax=-0.326$, $\sigma=0.097$ for the complex one.  The latter
is wider and fainter than the former, since the lowest-luminosity
stars of the clump are in this case partially included in the colour
bin used to compute the LF.  In any case, the \mimax\ values are 
0.17 -- 0.23 mag brighter than than those predicted for local stars.
The dispersion $\sigma=0.089$ can be compared the $\sigma=0.15$ value
found for LMC data by Udalski et al.\ (1998) and Stanek et al.\
(1998). It accounts for 2/3 of the observed value, which, as above
remarked, should include a dispersion in $A_I$ of at least 0.08~mag.

Which one of the above values of $\mimax$ should be preferred, is open
to discussion. In the case of Stanek et al.'s (1998) work, it is clear
that they sampled only the very red extremity of the clump in order to
define $\imax$. This case would better correspond to the
$\mimax=-0.384$ case in our simulations.

If we take this latter simulation as a reference, 
we conclude that the \mimax\ value for the LMC should be lower by
about $\Delta\mimax=-0.235$~mag with respect to the value obtained by
simulating the local clump (as shown in \reffig{fig_simconv}). If we
take at face value the mean $\iomax=17.835\pm0.008$ value obtained by
Stanek et al.\ (1998) for the LMC, and the reference value of
$\mimax=-0.209$ measured by {\em Hipparcos} data corrected for
Lutz-Kelker bias, we then derive a distance to this galaxy of
$\dmo=\iomax-\mimax-\Delta(\mimax)=18.279$. Uncertainties, however, are
not simply given by the uncertainty in photometry.  $\Delta\mimax$
values are model-dependent, and the numbers here used may be regarded
as a simple, first-order estimate for this quantity, uncertain to
about 0.1~mag.  Intrinsic errors in the \mimax\ determination and in
reddening (Udalski et al.\ 1998; Stanek et al.\ 1998) would add a
0.15~mag uncertainty to these numbers. 

Our conclusion, then, is that from this first simulation of the local
and LMC clump, the LMC should be located farther away by
$\Delta\dmo=+0.23$~mag with respect to the values obtained by Udalski
et al.\ (1998) and Stanek et al.\ (1998). This result is in
qualitative agreement with that obtained by Cole (1998), who 
from simple considerations about the age
and metallicity differences between local and LMC stars 
adopted a correction of $\Delta\dmo=+0.32$~mag.  Our results are
characterised by the use of more complex models, which 
recover many of the details of the clump defined by {\em Hipparcos}.

The LMC distance determination of $18.28\pm0.18$~mag is
compatible with the generally accepted value of about 
$18.5\pm0.1$~mag, found
by means of several different methods over the last years (e.g.\
Westerlund 1990; Panagia et al.\ 1991; Madore \& Freedman 1998;
Salaris \& Cassisi 1998; Oudmaijer, Groenewegen \& Schrijver 1998).

Our results can by no means be considered as a firm determination of
the LMC distance with the red giant clump method, because there remain
many uncertainties in the population syntheses to be solved, as there
are the mentioned assumptions about the age-metallicity-relation and
star formation history.

These uncertainties are absent for a population of single-aged stars
of identical composition as is present in globular
clusters. Therefore, we applied our method for testing purposes to the
galactic metal-rich disk cluster 47~Tuc. Kaluzny et al.\ (1998)
determine the extinction-corrected clump magnitude as $I_0^{\rm max}
=13.109\pm 0.026$. Using the {\em Hipparcos}-clump brightness as a
standard candle, they obtain a distance modulus of $(m-M)_0 =
13.34$. Our models, using a solar-scaled isochrone of 10 Gyr with
$Z=0.01$ and $Y=0.25$ predict $M_I^{\rm max}=-0.13$, with a dispersion
of only 0.014~mag, confirming the observed narrow brightness peak.
(An additional dispersion can be obtained if higher mass loss is
assumed.) However, this value for $M_I^{\rm max}$ has to be corrected
for two chemical pecularities: first, 47~Tuc shows $\alpha$-element
enhancement. Salaris \& Weiss (1998) showed that this leads to core
helium-burning models being 0.15~mag brighter than solar-scaled ones
at the same total metallicity. Second, Salaris \& Weiss (1998) argue
strongly that the helium content in 47~Tuc is close the solar value of
$\approx 0.27$, which again leads to brighter models ($-0.05$~mag),
such that the distance from the clump model is modified to $(m-M)_0 =
13.44$. For comparison, Salaris \& Weiss (1998), who determined the
age of 47~Tuc to $9.2\pm1.1$ Gyr, obtain $13.38\pm0.05$ from ZAHB
models. Within the uncertainties, this number agrees with both the
empirical and the theoretical clump distance. The larger difference
between the latter two values can be explained by the fact that 47~Tuc
has a lower metallicity than the average local clump star and
therefore is farther away than the standard-candle assumption of
Kaluzny et al.\ (1998) would predict. To close this discussion about
47~Tuc, Gratton et al.\ (1997) from {\em Hipparcos} subdwarf fitting
find $(m-M)_0=13.50$ and Reid (1998) with similar data and method
13.57. Of all values listed above, our clump distance agrees best with
these completely independent results.

Another source of error might be the transformations from theoretical
to observed quantities. We find, however, that our results change just
little if we use different tables of 
bolometric corrections. For instance, the $I$ bolometric corrections
derived from Yale tables (Green 1988) have a slightly lower dependence on
metallicity than those from Kurucz (1992). The use of these tables 
would cause an additional change of about $\Delta\dmo=-0.04$~mag in the LMC
distance modulus.

A further potential source of uncertainty is the dependence of \mimax\
with the helium content. Only to give an idea of the possible effects,
a change of $\Delta Y=+0.01$ would cause clump models of a given mass
and $Z$ to have clump magnitudes lower (i.e.\ brighter) 
by $-0.04$~mag.  Of course, any
of the theoretical predictions about the mean clump magnitude, depend
on the assumption that $Y$ increases with $Z$ according to a known
$Y(Z)$ relation. The relation adopted here represents a quite
conservative one, since present-day observations (see e.g.\ the
discussion in Pagel \& Portinari 1998) are not able to constrain the
$\diff Y/\diff Z$ ratio to the accuracy which would be desirable in
this kind of study.

\section{Final comments}
\label{sec_comments}

In this work we discuss the morphology of the red clump in composite
stellar populations and the possible origin of systematic errors in the
method of distance determination by means of these stars. These
systematic effects are such that \mimax\ depends only weakly on the \vi\
colour for a galaxy population (as the sample observed by {\em
Hipparcos}), but is systematically brighter for lower values of
the mean galaxy metallicity. This explains the low distance modulus
obtained for the LMC under the assumption that \mimax\ is independent of
age and metallicity (Udalski et al.\ 1998; Stanek et al.\ 1998), and
provides a value closer to the 
traditional one when the metallicity and age dependence suggested by
present models is considered.  This method of distance determinations
should therefore be used with care in galaxies where the history of
stellar formation and chemical enrichment has been very different from
the local galactic disk, as is the case for the Magellanic Clouds.

Our results also help to understand why the distance to M31 obtained
by means of this method (Stanek \& Garnavich 1998) is in agreement with
that obtained by other independent methods: it probably reflects the
fact that the mixture of stellar populations in the red clump of M31
[defined however in the $0.8<(\vi)<1.25$ colour interval] is similar to
the local disk one. The red clump method, in this particular case, may
work well with the assumption that the \mimax\ determined from {\em
Hipparcos} data could be compared to \imax\ observed in this galaxy.

Also in the case of galactic bulge stars, it may well be that the
distance determination by Paczy\'nski \& Stanek (1998) do not suffer
from significant systematic errors. This is for a different and more
subtle reason: In the case of stellar populations with 
metallicities from approximately solar to  twice solar (as for the
bulge), the trend of increasing \mimax\ with metallicity can be  
reversed if the helium content increases proportionally with $Z$. It
causes \mimax\ to be less sensitive to metallicity than for more
metal-poor populations. Such an effect can
be appreciated by comparing the $Z=0.019$ and 0.03 models in
\reffig{fig_zahb}. In this $Z$ interval, models of similar age differ
significantly in \vi\ colour, but not in absolute magnitude. This would
support the assumption that \mimax\ in Baade's Window stars is
similar to that given by {\em Hipparcos}, and would also explain the
absence on any trend in \imax\ with colour for them (Paczy\'nski \&
Stanek 1998).

On the other hand, the quest for a main revision of the calibration of
the extragalactic distance scale, that would be suggested by a short
distance to the LMC, is significantly weakened by the present results.
Indeed, they show that red clump stars cannot be considered as
standard candles. The assumption that \mimax\ is independent of the stellar
population can provide distance determinations accurate to only about
0.5~mag. Lower uncertainties are expected only in the case we have
additional information about the history of stellar formation and
metal enrichment of the galaxy to which the method is
applied. Therefore, this method of distance determination does not
improve over others currently used.

Another point of concern raised by the present work regards the
stellar models of clump stars. The fine structure observed in the blue
part of the red clump reflects directly the dependence of the core
mass at He ignition on the initial stellar mass. Our models
successfully predict this structure only because this dependence is
implicitly taken into account in all evolutionary sequences we
computed. However, models of red clump and horizontal branch stars
have often been computed under the assumption that the core mass at He
ignition is constant for all low-mass stars; this approximation has
been necessary because computing tracks up to the RGB tip is rather
time-consuming. Results of such calculations have been used by Cole
(1998), for instance. However, given the present results and the
quality of photometric data to come, such a simplification can no
longer be justified when the objective is the interpretation of the
red clump in composite stellar populations. Also, grids of
evolutionary tracks with mass resolution of 0.1~\Msun\ as here
adopted, or even lower, may be desirable in order to produce realistic
models of the red clump.

\section*{Acknowledgments}
Thanks are due to Y.K.\ Ng and P.\ Marigo for their useful comments,
and to B.\ Paczy\'nski for the debate which stimulated this work.
We made use of the Simbad database, mantained by the CDS, Strasbourg.
The work by L.\ Girardi is funded by the Alexander von
Humboldt-Stiftung.

\clearpage

\appendix
\section{Consideration of {\em Hipparcos} parallax errors in the LF}

\begin{figure}
\begin{center}
\psfig{file=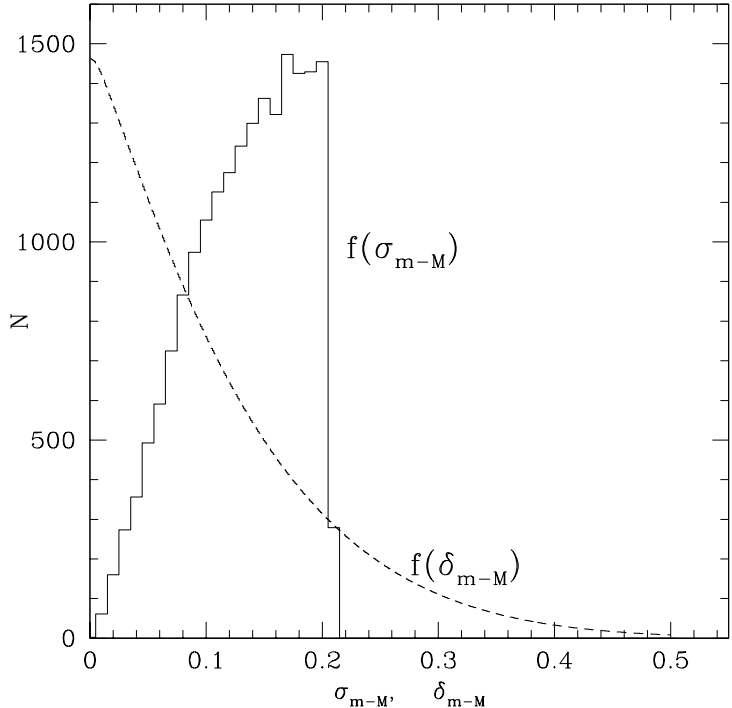,width=10cm}
        \caption{Distribution of standard errors in distance modulus,
$f(\sigmadm)$ for the stars in {\em Hipparcos} catalog with parallax
errors smaller than 10\% (continuous line). Assuming that errors in \dm\
are gaussian functions, the probability of having a \dm\ error of
$\deltadm$ is given by the dashed line.  }
\label{fig_ddm}
\end{center}
\end{figure} 

In \refsec{sec_model_mimax} 
it was argued that the theoretical LF for the local clump
has to be convolved with the error distribution of {\em Hipparcos}
parallaxes in order to allow a realistic comparison with observed data.
\reffig{fig_ddm} shows the distribution of standard errors
in distance modulus for the 19143 
stars with parallax known to better
than 10\% in {\em Hipparcos} catalog (using also additional quality
criteria similar to those defined in \refsec{sec_hippa}), 
or $f(\sigmadm)$.  A very similar distribution of errors is obtained
if we limit the sample to the clump stars. The mean error in distance
modulus for the complete sample is $\left<\sigmadm\right>=0.134$.

In order to simulate the effect of distance errors in our simulation,
we should convolve the theoretical LFs by a sample of gaussian curves,
each one of standard deviation $\sigmadm$, representing the complete
distribution of $\sigmadm$ values observed in the sample of {\em Hipparcos}
data. The resulting distribution of absolute errors, $f(\deltadm)$, 
is given by
\beqa
f(\deltadm) & = & \int_0^{\sigmadm^{\rm max}} 
	\frac{f(\sigmadm)}{\sigmadm\sqrt{2\pi}} \times \nonumber \\
	&  &
	\exp\left[ -\frac{(\dm)^2}{2\sigmadm^2} \right] 
	\diff\sigmadm	\;,
\label{eq_conv}
\eeqa
This function is also depicted in \reffig{fig_ddm} (only for
$\deltadm>0$). It includes the complete spectrum of distance errors
expected to be in the sample limited to 10\% parallax (standard)
errors.  An important point to be noticed is that this function is
already wider than the width of the theoretical distributions we have,
which have typically $\sigma\simeq0.08$~mag.

\label{lastpage}

\end{document}